# Low-cost vacuum compatible liquid cell for hard X-ray absorption spectroscopy

**C. Marini,**[a] **R. Boada,**[b] **J. Prieto Burgos,**[a] **N. Ramanan,**[a] **I. García Domínguez** [a], **L. Simonelli**[a]

[a] *ALBA Synchrotron, Cerdanyola del Vallès, 08290 Barcelona, Spain,*
[b] *Departament of Chemistry, Universitat Autònoma de Barcelona, Cerdanyola del Vallès, 08193 Barcelona, Spain*
  *E-mail*: cmarini@cells.es

ABSTRACT: We present the design, fabrication, and commissioning of a new liquid cell for X-ray absorption spectroscopy, that allows measurement in both transmission and fluorescence modes. The design consists of easily demountable and replaceable parts: body, kapton windows, silicon or viton o-rings, washers and a stopper screw. The pathlengths of the liquid chamber can be changed by simply substituting the elastomer o-ring, that can be cut from commercial silicon tubes, making the cell extremely customizable to overcome experimental constraints. The sealing of the liquid has been proved to be vacuum compatible. The compact and simple design makes the cell adjustable to cryogenic applications. Finally the advent of the 3d printing machines makes the cell potentially very competitive from a cost point of view (each unit is estimated less to be than 10 €)

## 1. Introduction

X-ray absorption spectroscopy (XAS) is a versatile technique for characterization and investigation of materials [1]. In contrast with diffraction methods or other optical spectroscopies (Raman, Infrared, Auger), XAS is an element specific probe which provides quantitative information about the local structure and electronic properties around an absorbing atom. Since it is based on the photoelectric effect, and on the consequential back-scattering of the ejected photoelectron waves from the neighboring atoms of the absorber, XAS only provides information about the short-range order. No assumption of periodicity or symmetry is done in its theoretical formulation [2], which means that XAS is equally applicable to condensed matter in all forms (solid crystalline or amorphous, bulk, liquid, and gas). The data are interpreted in terms of partial pair-distribution functions and the projected electron density of states, i.e. the type, number and distances of the surrounding atoms.

Liquids samples are investigated in several different fields of research. Chemical and biological reactions/processes mainly occur in condensed phases, i.e. solids and liquids. Indeed, many precursors for specific reactions are in liquid form [3] and, it is worth mentioning that some catalysts can show higher activity in liquid form [4]. Additionally, the high mobility of liquids has implications in electrochemistry, analytics, environmental science and pharmaceuticals [5-7]. Hence, it is imperative to have the possibility to probe local dynamics and geometry in liquids, since they directly determine their physical and chemical properties.

The potentiality of XAS in providing information on disordered matter (and in particular on liquids) has been widely demonstrated in the past years (a review is given in Ref. [8]). XAS can be measured in transmission and fluorescence modes. To measure in transmission mode with a reasonable signal-to-noise ratio, the absorption jump of the sample must be $\geq 0.1$. At the same time, in order to avoid compression of the spectral features, the total absorption of the sample needs to be controlled and limited. In the case of a liquid, these two factors lead to severe limitations in terms of the X-ray pathlength. Several technical solutions to measure liquids have been proposed for both hard and soft X-ray beamlines [9-11] in order to overcome the different experimental difficulties and constraints.

In Figure 1a we calculate (by using the XOP code [12]) the transmittance of X-rays through kapton films of several thicknesses as a function of the incident energy. It clearly shows that the cut off moves to higher energy for a larger thickness. This, in fact, ultimately determines the accessibility to the desired edges at lower energy. Another parameter to keep under control is the absorption length ($\lambda$) of the solvent, i.e. the thickness of the sample chamber, since it mainly determines the total absorption of the sample [13]. In Figure 1b, the absorption length of water, ethanol, methanol, and acetone are displayed as a function of energy.

Taking into account these general findings, here we try to overcome the thickness problem of the cell, proposing a new versatile liquid cell in polycarbonate, which is compatible with low vacuum (pressure $10^{-2}$ bars) for use in the hard X-ray range. Its extremely compact design, together with its vacuum compatibility, facilitates the adaptation of the design to cryogenic applications.

Finally, taking advantage of recent developments in 3d printing, this design results in extremely low production cost (0.60€/pc in polycarbonate and 8.30€/pc in PEEK), making it very competitive from an economical point of view. First commissioning results on chromium and selenium solutions are presented at room temperature and liquid nitrogen temperature.

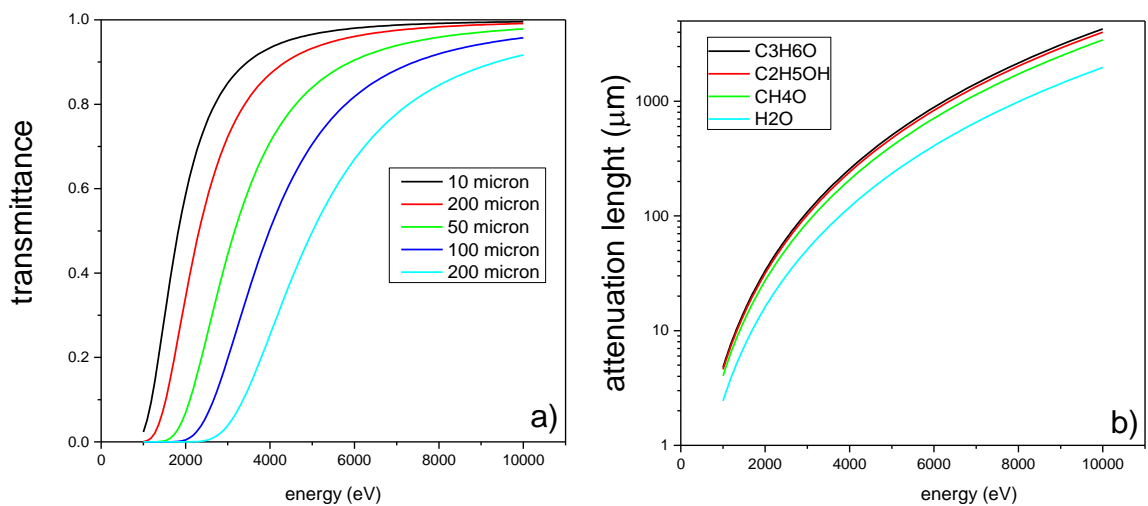

Figure 1: Transmittance of Kapton (panel a) and attenuation length of several solvent as a function of energy

## 2. Technical design

A technical drawing of the cell is shown in Figure 2. The cell is composed of three parts: body, sample enclosure (made of a succession of o-rings and windows), and a holding screw (M6 screw with a 4mm hole in the center). The body of the cell is made in polycarbonate, but the design can be adapted to Polyetheretherketone (PEEK). This materials choice makes use of the mechanical/corrosion resistance of these two plastics combined with their rather good thermal properties [14].

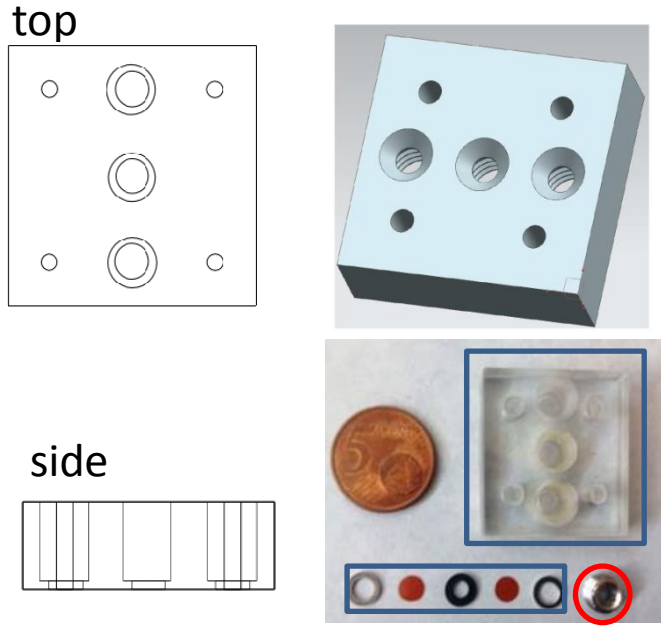

Figure 2: (a) Schematic view of the cell for X-ray absorption investigations of liquids (top and side view drawing) and (b) experimental set-up used at CLAESS. The three essential parts (body, sample assembly and fixing screw) are marked by rectangles and circle

The cell has been designed for transmission measurements but it is also compatible with fluorescence measurements. Special care has been adopted in the design of the top wall, where an angular aperture of 90° has been chosen to permit the collection of spectra from the cell in fluorescence mode without shadow effects from the body. Due to their chemical composition, the use of polycarbonate or PEEK minimizes the contribution of additional lines to the fluorescence spectra of the liquid sample. The body of the cell hosts three liquid chambers (that fits the cryostat setup of CLÆSS beamline [15]), thus permitting the simultaneous loading of three liquid samples, thus reducing the beamtime lost during sample changes. The sample packaging consists in an aluminium washer followed by a first kapton window and an elastomer o-ring, whose thickness can be chosen according to the required X-ray path. The liquid sample is loaded by dropping the liquid into the o-ring. The compartment is then sealed with another kapton window, followed by another metallic washer. We chose Aluminium because, in the case of fluorescence measurements, its fluorescence line (at 1486 eV) is completely screened by the body of the cell. Finally, the holding screw presses the assembly against the wall of the cell body. O-rings of several thicknesses are commercially available but, in order to customize the cell to specific experimental constraints, it is possible to replace the elastomer o-ring with a piece of commercial silicon tube (#16 from Watson Marlon [16]) cut to the desired length up to a maximum of 5 mm liquid path. For commissioning purposes, the sample thickness has been changed from 1 mm to 2 mm. We tested the resistance of the kapton window to stress after loading the liquid by performing several pumping/venting cycles. No leakage and/or degradation of vacuum performance of the cell have been observed.

## 3. Data and Discussion

In the following we report about the commissioning of the cell. We have performed XAS measurements on Cr and Se salts dissolved in water at Cr K-edge (5989 eV) and Se K-edge (12658 eV) respectively. The choice of Cr and Se is based on their role in biological metabolism as well as environmental pollution.

Cr is recognized as an essential nutritional element for human health due to its role in glucose metabolism [17]. However, ingested Cr above 0.05 mg $L^{-1}$ may have serious consequences including tumours, ulcers, and cancer [18, 19]. In addition, Cr(VI), has been found to be more toxic than Cr(III) to various microbiota [20]. Se has chemical properties similar to sulfur, but slight differences can lead to altered tertiary structure and dysfunction of proteins and enzymes [21].

Elevated Cr and Se concentration in drainage water and shallow ground waters are known to cause bioaccumulation of Se in plants and animals above toxicity levels. This is the reason why several XAS investigations have been performed on aqueous solutions containing these two elements [22, 24].

### 3.1 Experimental details

The X-ray absorption experiments presented here have been performed at CLAESS beamline at the national Spanish light source facility ALBA [15]. Transmission EXAFS measurements at Cr and Se K-edges were performed in Q-EXAFS mode, using a Si(111) and Si(311) double-crystal monochromator, respectively. By means of a vertically focusing mirror, the beam has been focused into the sample down to 500x500 $\mu m^2$, although a bigger beam spot is compatible with the cell aperture (4 mm diameter). The incoming and outgoing photon fluxes (respect to the sample) were measured by ionization chambers filled with appropriate mixtures of $N_2$ and Kr gases. The cell has been inserted into the liquid-nitrogen cryostat of the beamline, and the temperature was varied between 77 and 300 K, controlled by a feedback loop electric heater.

Although reasonable data quality was achieved with a single scan (about 3 minutes), in order to check reproducibility several repeats have been collected on the same liquid.

At low energy we used a 350 mM solution of $Cr(NO_3)_3(H_2O)_9$ and 116 mM solution of $K_2Cr_2O_7$ as representative of $Cr^{III}$ and $Cr^{VI}$ samples respectively, while at high energy, 200 mM solutions of $Na_2SeO_3$ and $Na_2SeO_4$ was chosen for $Se^{IV}$ and $Se^{VI}$ standards. Different thicknesses of the liquid cell chamber have been used to keep the total absorption effect on the spectra under control. At Cr and Se K edges, the x-ray path lengths were 1 mm and 2 mm respectively, giving rise to a jump of ~ 1.2 (Cr K edge) and 0.92 (Se K edge) with a total absorption below 1.5.

A metal foil of Cr and a pellet of metal Se have been measured to calibrate correctly the energy scale of the monochromator. Data reduction and data analysis were performed according to standard procedures using DEMETER software package [25].

## 3.2 Spectroscopic data

Cr K edge XANES spectra are shown in panel a) of Figure 3. When Cr is in non-centrosymmetric tetrahedral Cr(VI) configuration, a prominent pre-edge peak occurs at 5992 eV, due to the 1s to 3d transition, enormously enhanced by the empty 3d-state of Chromium (i.e. $Cr^{VI}$ is $3d^0$). This transition is forbidden by symmetry rules in octahedral $Cr^{III}$ configuration that is centrosymmetric. Nevertheless, small pre-edge features are present in $Cr^{III}$ at 5989 and 5992.5 eV, which are due to 1s to 3d ($t_{2g}$) and 1s to 3d ($e_g$) electronic transitions, respectively. The lower intensities of these features can be associated to the stability of the $3d^3$ electronic configuration [26]. The 3 eV energy separation between the $Cr^{III}$ pre-edge features is in agreement with the 2-3 eV octahedral crystal field splitting between the $t_{2g}$ and $e_g$ levels measured for several $Cr^{III}$ compounds in octahedral coordination [27]. The study of the intensities of the pre-peaks in Cr K edge has been proposed to quantify and distinguish between the $Cr^{VI}$/total Cr content in Cr-contaminated soil and water samples [23].

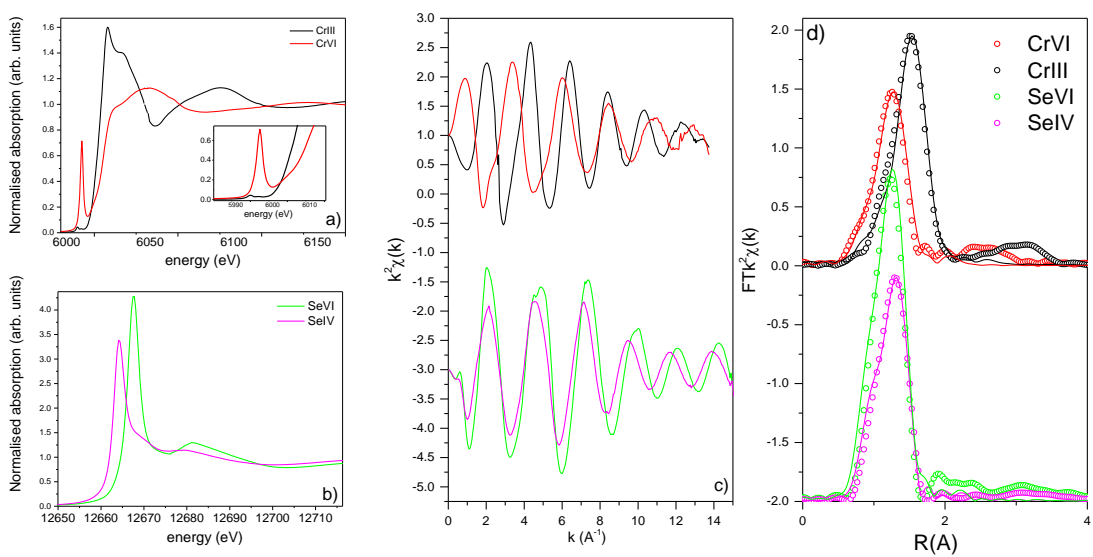

Figure 3: XANES spectra of Cr (a) and Se (b) aqueous solutions. EXAFS data (b), Fourier Transform and first shell fit (d) of the spectra collected at the two different energies.

Se K edge XANES spectra are shown in panel b) of Figure 3. Irrespective of the specific valence of Se, both spectra are characterized by a strong white line (located at 12664 and 12667 eV for $Se^{VI}$ and $Se^{VI}$ respectively), corresponding to 1s to 4p dipole allowed transition. The white line in $Se^{VI}$ standard is more pronounced than in $Se^{IV}$. As in the Cr K edge spectra, different valences of Se determine different local arrangements around the absorber site: $Se^{IV}$ assumes a trigonal pyramidal configuration (as $As^{III}$) [28], whereas $Se^{VI}$ assumes a tetrahedral configuration (as $Cr^{VI}$ and $As^V$).

EXAFS spectra at both Cr and Se K edges are reported in panel c) of Figure 3. The EXAFS signal appears very simple with one main frequency wave, as usually observed in disordered systems. Beside such simplicity, the different local arrangements of the Cr and Se atoms result in different oscillation patterns, characterized by different frequencies, as confirmed by the Fourier Transforms in the k-range 2.5-12.5Å$^{-1}$ reported in panel d) of Figure

3. As evident, the main peak is displaced at a different R position corresponding to the metal-O distances in the octahedron/tetrahedron (Cr) or in the pyramid/tetrahedron (Se).

In order to quantify these findings, we fit the data according to the standard EXAFS equations [29]. Structural model used in the analysis has been derived from the undiluted salts. $\Delta k$ and $\Delta R$ windows for the fits are respectively 10 Å$^{-1}$ and 1.5 Å. For the first shell we fixed the coordination number to the one of the corresponding solid phase, and we use one distance and 1 Debye Waller factor ($\sigma^2$) to parametrize the EXAFS pattern. Best fit curves are reported in the same panel. Results of the analysis are summarized in table I.

The tetrahedral configuration is characterized by the highest bond strengths (lower Debye Waller factor) and the shortest metal-oxygen distances. These results are consistent with literature [23, 30].

Table 1: EXAFS analysis results at room temperature.

|  | N | R (Å) | $\sigma^2$(Å$^{-2}$) |
|---|---|---|---|
| CrIII sol | 6 | 1.96 | 0.004 |
| CrVI sol | 4 | 1.72 | 0.006 |
| SeIV sol | 3 | 1.69 | 0.004 |
| SeVI sol | 4 | 1.652 | 0.003 |

In order to demonstrate the compatibility of the cell with cryogenic applications, the same 200 mM solution of Na$_2$SeO$_4$ (Se$^{IV}$ standard) was loaded into the cell, deep quenched at 77 K (by immersing the cell in liquid nitrogen) and fast cooled down to 77 K into the cryostat. The XANES spectra at 300 K and 77 K are shown in Figure 4.

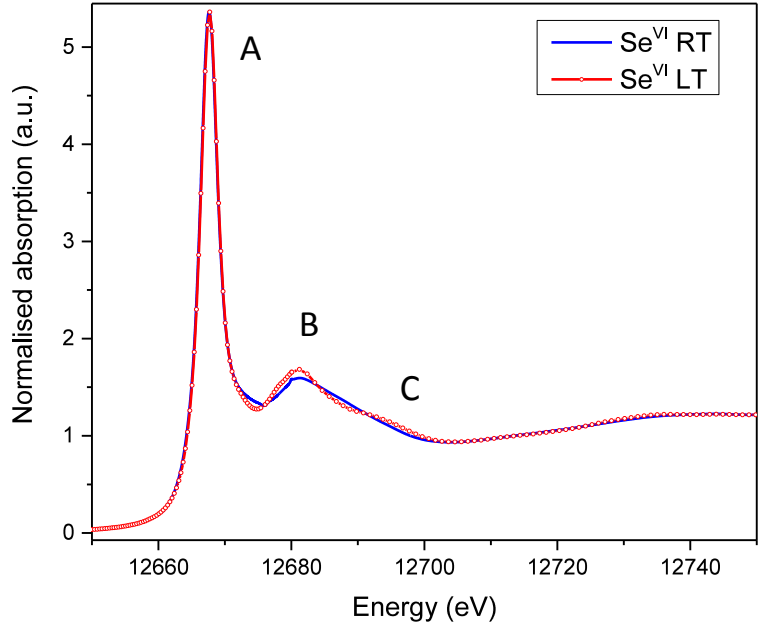

Figure 4: XANES spectra of aqueous Na$_2$SeO$_4$ (SeIV standard) solution at room temperature and liquid nitrogen temperature after quenching the sample.

While the edge position and intensity of the 'white' line (labelled as A) are approximately the same, a clear difference between the two spectra can be observed in the two features located around 12665 eV (labelled as B) and 12675 eV (labelled as C) respectively: both features gain intensity and become narrower as the temperature is decreased. These findings can be explained bearing in mind that quenching the sample means acting on the configurational disorder of the system, and different degrees of disorder corresponds to different intensities in the XANES features.

## 4. Conclusions and outlook

A novel setup for temperature-dependent XAS studies on of liquid samples is presented. The proposed liquid cell allows measurements in both transmission and fluorescence modes. We have demonstrated its correct working using aqueous solutions containing Cr and Se salts. The sample packing and sealing system is extremely simple and customizable. Moreover, the cell is vacuum compatible, extending its use to low energies and cryogenic applications. To demonstrate this, XAS spectra were collected at room and liquid nitrogen temperatures, after deep quenching of the samples.

Finally, because of its simple and compact design, 3d printing machine allowed the rapid creation of its constituent parts, with accuracy and cheap prices, thus making the diffusion and use of the cell very competitive.